\documentclass[fleqn,10pt]{wlscirep}
\usepackage[utf8]{inputenc}
\usepackage[T1]{fontenc}
\usepackage{comment} 
\usepackage{float}   
\usepackage{multicol} 

\title{Artificial Neural Network for Estimation of Physical Parameters of Sea Water using LiDAR Waveforms  
}

\author[1, 2, 3,*]{Saad Ahmed Jamal}

\affil[1]{
MED – Mediterranean Institute for Agriculture, Environment and Development 
Instituto de Investigacao e Formacao Avancada, Universidade de Evora, Apartado 94, 7002-544 Evora, Portugal.
}
\affil[2]{Department of Geoinformatics - Z\_GIS, Paris Lodron University of Salzburg, 
Austria}
\affil[3]{Department of Computer Science, University of South Brittany, 
France}

\affil[*]{saad.jamal@uevora.pt}

\begin{abstract}

Light Detection and Ranging (LiDAR) are fast emerging sensors in Earth Observation. The potential of Full Waveform LiDAR (FWL) is much greater than just height estimation and 3D reconstruction. Most LiDAR software works on point clouds by utilizing the maximum value within the waveform while the shape is left underexplored. The overall shape of the signal provides important information about the properties of the water body. Existing techniques in LiDAR data analysis include depth estimation through inverse modelling and regression of logarithmic intensity and depth for approximating the attenuation coefficient. However, these approaches have their limitations. Additionally, no established modeling method existed for predicting bottom reflectance. This research proposed a novel solution based on neural networks for parameter estimation in LiDAR data analysis. The proposed model successfully learned the solution by using simulated data to overcome the LiDAR data scarcity issue and by leveraging one dimensional convolutions in deep learning model. It made predictions of parameters, which include depth, attenuation coefficient, and bottom reflectance, with r2 scores of 0.64, 0.95 and 0.35, respectively. The model's performance was validated by testing it on multiple test sets, including real LiDAR data. More data availability would enable better performance and reliability of such models.

\end{abstract}
\begin{document}

\flushbottom
\maketitle

\thispagestyle{empty}


\section*{Introduction}



Light Detection and Ranging (LiDAR) is an active Remote Sensing (RS) technology used for high-resolution mapping. It utilizes laser beams to measure distances and create detailed three-dimensional representations of objects and environments. It uses laser beams to provide high-resolution spatial data. Modern sensors have a resolution of a few centimeters. A typical LiDAR instrument consists of a laser, a scanner and a specialized Global Positioning Systems (GPS) receiver. The principle of LiDAR is to measure the time of flight taken by the pulse to reach the receiver from the transmitter after reflecting from the object. The analysis of the back-scattered signal (in particular its maxima), also called waveform, enables the derivation of 3D point clouds related to encountered objects  \cite{1_ReviewBALTSAVIAS199983, 2_Review0_MALLET20091, 3_Review3_Indian_Mehendale_2020}. It is to be noted that LiDAR can only measure distance using time measurements given by equation \ref{eq1}.
\begin{equation}\label{eq1}
D = c  (\Delta t/2) 
\end{equation}
where \textit{D} is the distance of the object, c is the speed of light, and \textit{$\Delta$t} is the time required to travel by the laser. LiDAR scanning mechanism is mainly classified into four types: (i) Opto-mechanical, (ii) electro-mechanical, (iii) micro-electro-mechanical systems (MEMS), (iv) solid-state. The electro-mechanical scanning is mostly used in LIDAR nowadays \cite{4_Raj_2020}. Efforts are being made to use Solid State Scanning more often due to its potential robustness, field of view (FOV), and scanning rate potential. The resolution or minimum detectable object essentially depends on the object's reflectivity. Several other factors, such as atmospheric conditions, background irradiation, type of target, noise level, laser wavelength, aperture, detector sensitivity, angle of inclination of surface, and shape of target \cite{5_Review2BALTSAVIAS1999199}, contribute to detectability.

A new LiDAR simulator for waters (WALID) produces waveforms for flooded (water-covered) areas similar to those coming from an actual LiDAR sensor \cite{6_WALID}. LASER wavelength, generated by this simulator, is from 0.3 to 1.5 micrometres, corresponding to the ultraviolet to infrared range. GLAS/ICE-SAT is a satellite that provides altimeter data through elevation measurements \cite{7_Baghdadi_2011}. For the infrared region, GLAS system characteristics were used for the waveform simulation while for the visible range in WALID. HawkEye LiDAR is an airborne sensor for bathymetry \cite{8_CHUST2010200}. It can measure sea bottom elevation in low water depth areas. GLAS and Hawkeye were used to compare the observed waveform from the simulated waveform using signal-to-noise ratio and homothetic transformation. The novelty of the WALID simulator model lies in considering the physical properties of water, such as the transfer of energy in the form of electromagnetic radiation through a medium (laws of radiative transfer). There are several other simulators available. Virtual environments such as computer games were used as simulators. GTA-V game was used to generate a large amount of realistic training data in a virtual environment \cite{9_GTA-V_8462926}. Squeezeseg, a network based on the convolutional neural network (CNN), was used for semantic segmentation and point-wise classification of the simulated data.

LiDAR is rapidly becoming popular for ground mapping. It has proved reliable for various oceanography, forestry, hazard monitoring and geology applications. The geographical Information Systems (GIS) world has known about this sensor for a decade, but other systems are now adopting it for multiple uses. For example, LiDAR is now being used in smartphones for better-quality photography. The new generation iPhone 12 Pro contains a LiDAR sensor, which provides 3D animations \cite{10_TAVANI2022103969}. Modern sensors are allowing high-resolution mapping on a large scale. Topo-Bathymetric LiDAR is replacing previous means of seismic and bathymetric surveys near the ocean water and land interface, which were extremely expensive and cumbersome. Unmanned Aerial Vehicle (UAVs) equipped with LiDAR allows repetitive surveys now used for temporal analysis\cite{11_Hameed_https://doi.org/10.1155/2022/6430120}. Uses of LIDAR in RS include terrain mapping, digital elevation modeling, forest inventory, biomass estimation, coastline, beach mapping, floodplain mapping, river channel mapping, urban planning, building height determination, mining and mineral exploration, archaeological site mapping, snow mapping, oceanography and coastal zone management, precision agriculture, and crop mapping and transportation infrastructure and asset management\cite{3_Review3_Indian_Mehendale_2020,12_mallet:hal-02384727}. LiDAR's technology limitations include high costs for data acquisition, and maintenance, large data volumes, complex data processing, potential gaps, limited footprint, and wavelength range \cite{13_BELAND2019117484}. It is a line-of-sight technology. LiDAR waves are also affected by weather and the objects they interact with.

LidarNet, developed by Andreas et al. \cite{14_Simulator2}, regenerates the waveform datasets for specific parameters. According to their statement, they have achieved 99\% accuracy by employing a model based on CNN. The authors trained a classifier on simulated data and achieved favorable results for validating unseen real data. Andersen et al. \cite{15_ANDERSEN2005441} created a regression model for forest canopy fuel estimation. A strong correlation was discovered for all parameters between metrics derived from LiDAR data and fuel estimates obtained through field-based measurements. The research showed that LiDAR-based fuel prediction models can be used to develop maps of canopy fuel parameters over forest areas. Koetz et al. \cite{16_1576688} used LiDAR waveform for the estimation of the biophysical parameters of the forest. The forest canopy structure was determined by deriving parameters such as leaf area index, vertical crown extension, maximum tree height, and fractional cover. Two data sets, synthetic and real, were used to evaluate the new method's accuracy. The data included laser waveforms and associated forest canopy information, which was used to assess the performance of the proposed approach. A synthetic data set was created by simulating the waveform response of 100 simulated forest areas using the LiDAR waveform model described earlier. A real-world dataset was collected in the Eastern Overpass Valley, within the Swiss National Park. Extracting information from the LiDAR waveform model was based on a Lookup Table (LUT) method. This approach involves building the LUT and selecting the appropriate solution matching a specific measurement. 
\begin{equation}\label{eq2}
X^2 = \sum\limits_{i=1}^{n_{bin}}\,(w_{ref}^i-w_{sim}^i) 
\end{equation}
The result of the model inversion was determined by reducing the value of the merit function ($X^2$), which measures the difference between the reference waveform $(w_{ref}^i)$ obtained from the laser system and the simulated waveform ($w_{sim}^i$) stored in the LUT. The simulated waveforms were scaled based on their highest peak, making them consistent with the recorded signal using equation \ref{eq2}.
It is also common to use observations of the physical system to solve the inverse problem, that is, to learn about the values of parameters within the model. This process is often called calibration. The main goal of calibration is usually to improve the predictive performance of the simulator \cite{17_Brynjarsdottir_2014}.

Artificial Neural Networks (ANN) have been used for the estimation of abstract parameters in other domains. ANN was used by Calder et al. \cite{18_Calder_n_Mac_as_2000} for geophysical parameter estimation, including formation resistivity and layer thickness. Seismic waveform data was used as input data for the feed-forward neural network. Such deep learning frameworks have proven to be useful for the analysis of LiDAR data. Algorithms can act as a set of mathematical transformations \cite{19_8465968}. For problems such as building detection, mathematical computations alone are not adequate to model the physical properties of a problem; in such cases, raw LiDAR data has to be augmented with features coming from the physical interpretation of data.
Convolutional Neural Network (CNN) has been the most effective deep-learning approach for dealing with image data \cite{20_MALLET2011S71}. The temporal CNN developed by Pelletier et al. \cite{21_CharlottePelletier_rs11050523} introduced an effective deep-learning approach for classifying time series data. This technique utilizes temporal convolutions (convolutions in the temporal domain) to learn both temporal and spectral features automatically. This led to better results than the results of the Recurrent Neural Network (RNN) and Random Forest (RF), with a margin of about 1-3\% overall accuracy. However, the lack of integration of spatial domain causes salt and pepper noise in the results. The architecture comprises three convolutional layers with 64 units, one dense layer with 256 units, and one softmax layer. The paper also demonstrates the effect of filter size, batch size, model depth, pooling, and spectral and temporal dimensional guidance. Hybrid methods have proven more effective when dealing with time series data. Hybrid methods combine quantitative time series models with deep learning \cite{22_HAMMAD2023100848}.

Several techniques are used for bathymetric mapping. Spectral methods can measure depths up to 2 to 2.5m with a mean error of 15–20cm. Structure from motion, which uses stereo imaging for 3D modeling, is also used to measure heights. However, it is ineffective for measuring through water as it is effective only up to 2m inside water. Coastal bathymetric with airborne LiDAR uses green laser 532nm, one of the least absorbed wavelengths \cite{23_Letard_2022}. It can measure depths up to 40–50m. However, the cost of such a survey is costly with prices as high as 1000 \texteuro/km\textsuperscript{2}. Multibeam Sonar is the most useful method for measuring deep water courses. With a range of up to 100 to 4000m, it uses sound waves to propagate through the water to produce a bathymetric map \cite{24_Conference_5151853}. The limitation of this technique includes lesser efficiency for bathymetry at shallow water depth.

Topo-bathymetric through airborne LiDAR, though, cannot go hundreds of meters into the water, yet it is a useful technique for high-resolution and high-precision surveys. Awadallah et al. \cite{25_NewTest_rs15010263} did a comparison among different bathymetric sensors. They concluded that these sensors could provide high-quality river geometry representation. It could penetrate up to 80m into water. The penetration depth depends on several factors, such as the laser's wavelength, the system's power, the seafloor's reflectivity, and the water quality, such as turbidity, viscosity, surface roughness, and suspended particles in water. Many of the water's physical parameters are unknown, making it difficult to calculate the empirical penetration depth \cite{26_LAGUE202025}.

The backscatter signals over water for full waveform consist of three parts. i) Surface echo when the wave strikes the surface, ii) an exponential attenuation of the signal in the water column because of absorption by water and suspended particles, and iii) bottom echo that corresponds to bed reflectivity. Depending on bottom topography, the shape of the backscattering echo differs \cite{27_survey4_s23010292}. Topo-bathymetric LiDAR has the potential to replace numerical models based on channel cross-sections that are used for large-scale hydraulic modeling. A similar network was used to classify coastal and estuarine ecosystems \cite{28_Letard9554262}. Awadallah et al. \cite{25_NewTest_rs15010263} also analyzed the discrepancies in altitude among the bathymetric, MBES and TLS point clouds. It also assessed the differences in altitude within the bathymetric LiDAR point clouds. It connected these variations to river characteristics such as depth, bank inclination, sudden changes in altitude, and turbulent areas. The bathymetric measurements acquired in the autumn season tend to have more errors with less penetration within water than the ones acquired in winter \cite{29_east_pak}.
This seasonality effect is due to higher sediment concentration in autumn. Attenuation is the reduction in the strength of a signal, wave, or other form of energy as it travels through an environment. Attenuation can occur due to several factors such as absorption, scattering, reflection, and dispersion. Based on the backscattering of the LiDAR beam, Yang et al. \cite{30_attenuation_photonics9100713} developed equation \ref{eq3} for calculating the attenuation coefficient (kd) of water. The attenuation coefficient was measured as 1.343 for their experiments.
\begin{equation} \label{eq3}
    \ln(E_i) = - C ( z + 0.2)+\ln (\eta E_0)
\end{equation}
where \textit{C} is the attenuation coefficient of water, \textit{E0} and \textit{Ei} is the sent and received power of source light, and \textit{z} is the measuring distance between the reflector and light source.

This study is the first attempt to predict parametric values by learning patterns through the LiDAR waveform. The objectives that were pursued during the research were:  (i) create a large amount of simulated data, (ii) design a deep neural network using simulated data, and (iii) evaluate performance on real data. The current knowledge domain lacks the accuracy to calculate or retrieve these parameters. Out of the three parameters under study, depth, attenuation coefficient, and bottom reflectance, no current modelling method exists for extracting this information. Existing techniques in LiDAR data analysis include depth estimation through inverse modelling and regression of logarithmic intensity and depth for approximating the attenuation coefficient. However, these approaches have their limitations. Depth estimation through inverse modelling provides only approximate values. It does not account for variations in surface properties. At the same time, the regression approach for the attenuation coefficient can only generalize a value through several data points, which lacks accuracy and may lead to significant errors in estimation. Additionally, there existed no established modeling method for predicting bottom reflectance. This research proposed a deep learning model with the help of convolutional neural networks which can predict the three parameters. The model is more accurate and applicable and requires less computational power than previously existing solutions. This introduces a new dimension to LiDAR data analysis, enhancing its usefulness and productivity for bathymetric applications.

\subsection*{Study Area}

\begin{figure}[H]
    \centering
    \includegraphics[width=1\textwidth]{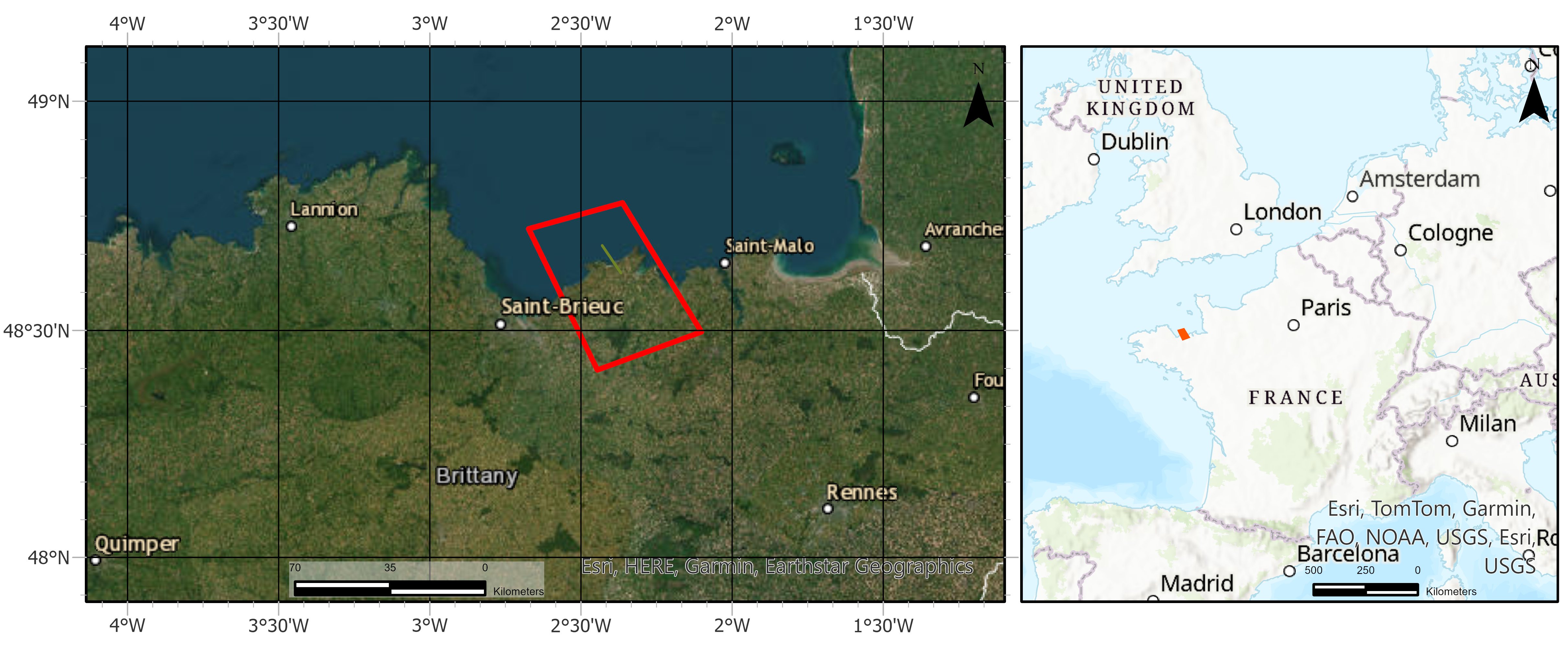}
    \caption{The study area map shows the geographical location of the data collection area generated using \href{https://www.esri.com/en-us/arcgis/products/arcgis-pro/overview}{ArcGIS Pro 2.6.0} a license provided by the University of Salzburg.}
    \label{fig1:StudyArea}
\end{figure}

The study area is Frehel, a commune located in Bretagne, the western region of France. It is situated along the Emerald Coast along the English Channel. Frehel has cliffs, sandy beaches, rugged landscapes, and few inhabitants. Its geographical coordinates are approximately 48.6308° N (latitude) and -2.3899° W (longitude). Fréhel experiences an oceanic climate, characterized by mild temperatures and moderate yearly rainfall. Summers are relatively cool, with average temperatures between 15–20°C, while winters are mild, rarely dipping below freezing. The proximity to the English Channel ensures high humidity and frequent winds, which contribute to the dynamic coastal environment. The topography of Fréhel is diverse. The area features rugged coastal cliffs, such as the famous Cap Fréhel, which rise to approximately 70 meters above sea level. These cliffs provide panoramic views of the English Channel and are surrounded by heathlands and moorlands rich in biodiversity. Fréhel's combination of unique topography, diverse soils, and coastal climate makes it an important area for ecological and environmental studies, particularly in understanding coastal dynamics, biodiversity, and the impacts of climate change. Figure \ref{fig1:StudyArea} shows a satellite view of the study area.

\section*{Methodology}

The aim was to determine the parameters of the target water body from waveform mainly reference water depth, attenuation coefficient for the water, and bottom reflectance. In this research, we primarily used the WALID simulator \cite{6_WALID} to generate many sample waveforms. The simulator was accessed through an anaconda interface. Simulated data was generated for the interaction of LiDAR signal over the water body for the purpose of bathymetric analysis. This was made by defining a set of input parameters, which are described later in Table \ref{tab1:InputParameters}. The simulated data was used to feed the neural network for making the prediction of the required parameters. The reason for using simulated data was to train a deep neural network with sufficient amount of input data. Because it was not possible to gather an ample amount of real LiDAR data to train an efficient neural network. Real data from actual LiDAR surveys is not readily available and is costly. The strategy chosen was to train the network on simulated data and use the real LiDAR data for validation purposes. Also, in general, the larger the amount of data, the better weights are learned and the better the neural network. So, the option to train using simulated data seemed to be effective, as an ample amount of simulated data could be generated using GPUs.

\begin{figure}[H]
    \centering
    \includegraphics[width=1\textwidth]{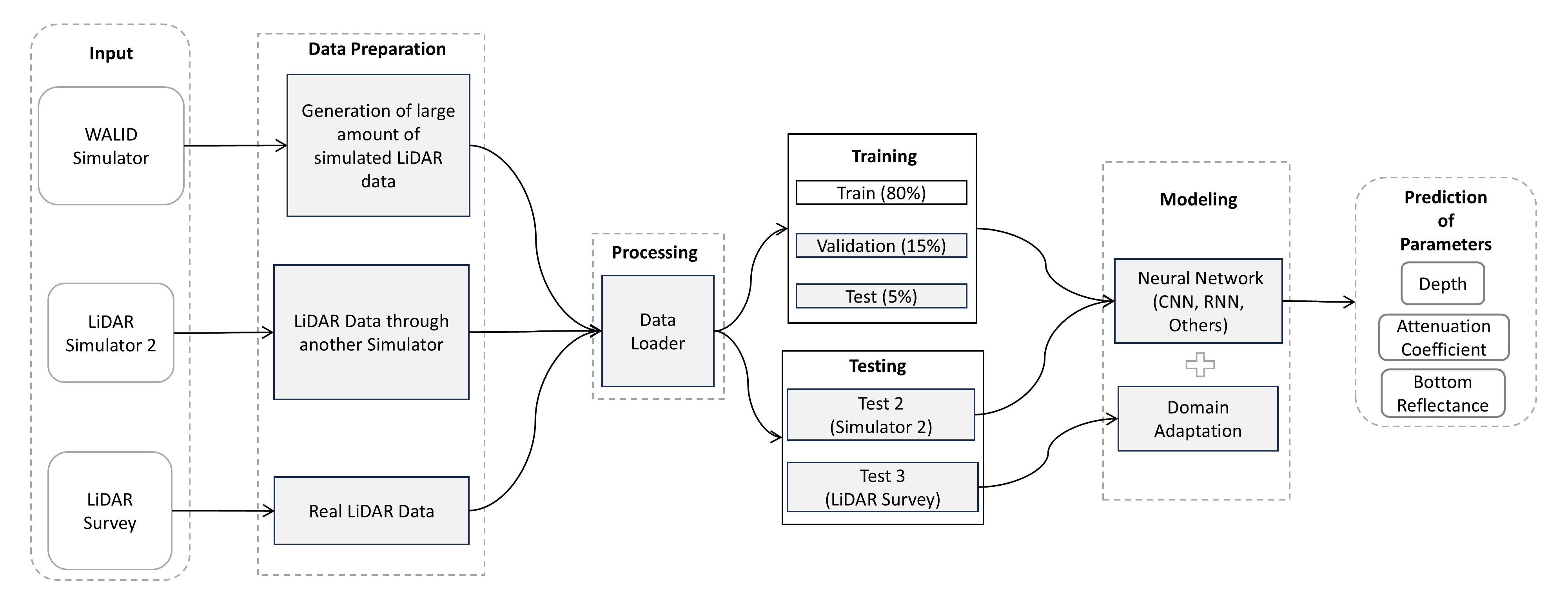}
    \caption{Methodological Flowchart}
    \label{fig2:MethdologyFlowchart}
\end{figure}

A common question often addressed during the research was why deep learning should be used instead of inverse, mathematical, or other models. This is because no such model exists for these parameters except for one parameter, depth. Depth can be calculated using inverse modeling, but that is computationally expensive. So, deep learning is more useful in such cases. Scalability is one of the main advantages of deep learning models in that they show good performance with large amounts of data. The deep network was coded using the TensorFlow library in Python. Figure \ref{fig2:MethdologyFlowchart} presents an abstraction of a methodological flowchart.

The network was first trained to learn and approximate the relation between the optical parameters and the backscattering of the signal from the water. This will result in a data-driven model capable of mapping the relationships among the parameters and waveform. The second task is to invert this model to solve the inverse problem to produce parameters that honor the known waveform and reproduce the water characteristics. Data from WALID Simulator was first used for training in the pipeline as shown in Figure \ref{fig2:MethdologyFlowchart} and later second simulator data and LiDAR Survey data was used for during testing phase. Also, simulated cases allow for a better evaluation of the accuracy of the technique used to estimate the missing parameters. The total power received by LiDAR is the sum of these parameters with the addition of factors equation \ref{eq4}.
\begin{equation}\label{eq4}
    P_T (t) = P_s(t) + P_c(t) + P_b(t) + P_{bg} (t) + P_n (t) 
\end{equation}
where $ P_T (t) $ is the total power received, $ P_s(t) $ is the power returned by the water surface, $ P_c(t) $ is the power returned by the water column, $ P_b(t) $ is the power returned by the bottom, $ P_{bg} (t) $ is the background power returned by the air column, Pn (t) is the detector noise power and t is the time scale. 
A time series is a collection of data points measured at regular intervals over time.

\subsection*{Generating large input data}
First, it was essential to produce a large amount of training data. The WALID simulator was used with suitable parameters to generate many datasets. Parameters were selected according to the Titan sensor to simulate a waveform similar to the real LiDAR sensor, recorded by the sensor for the green wavelength. The parameters were amplitude (scales the whole waveform), $imp_type$ (extreme dist, GevII), a data sampling interval, the waveform's base intensity, noise, and water surface echo reference intensity. Apart from these, medium parameters were also selected such as reference intensity of water decay ($I_w$), the attenuation coefficient of water (kd), bottom reflectance $(I_ref)$, depth of water, and maximum depth of the medium.

On a local desktop computer, generating substantial data (exceeding 100,000 waveforms) and processing the resulting datasets was not possible. The cluster, a high-performance computing facility, was used for this purpose. This cluster had Graphical Processing Unit (GPU) resources supporting up to 1152 GB of Graphical Memory with individual GPU units of 256 GB. The cluster was accessed through a Linux command line interface. Waveform generator code was uploaded onto the cluster. Subtle changes were required to run the code on the cluster compared to the local version. To generate one million waveforms on the cluster, the simulation took 14.2 hours. The neural network could be sufficiently trained with this number of training data. Figure 3 shows a few random samples of simulated waveforms where the y-axis shows the intensity of the received signal.

\begin{figure}[H]
    \centering
    \includegraphics[width=1\textwidth]{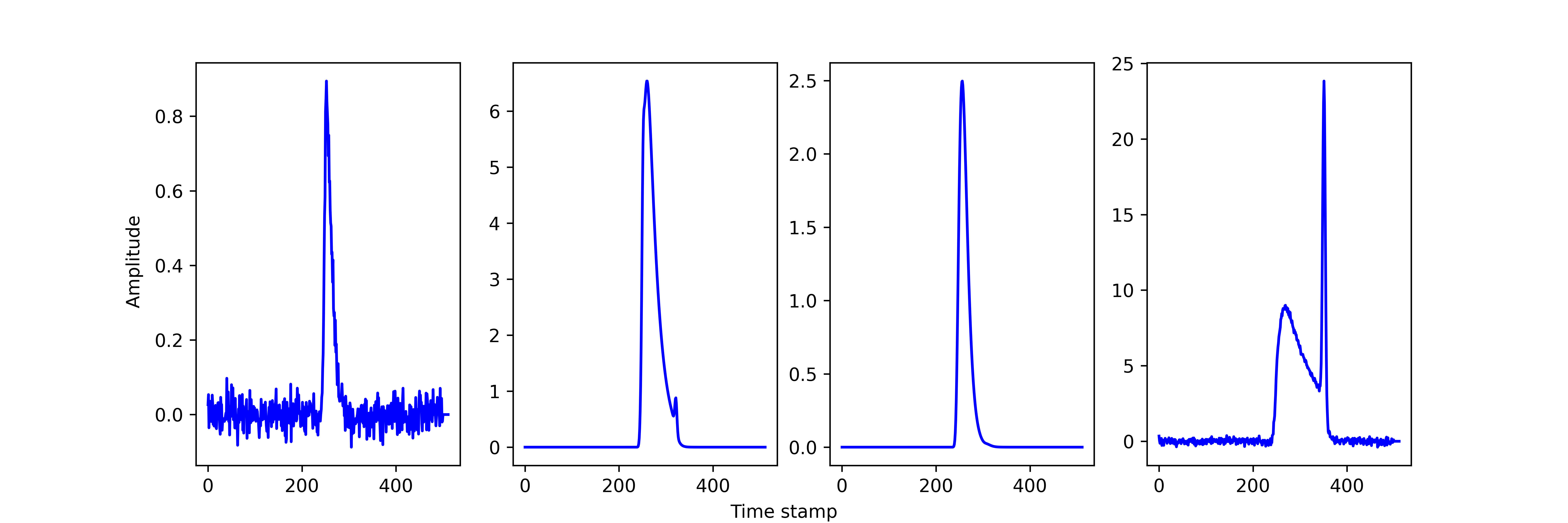}
    \caption{Sample plots of simulated Waveforms generated through WALID}
    \label{fig3:SampleWaveforms}
\end{figure}

\subsection*{Design of the model}

Then came the design of the most appropriate model that minimizes the loss and error in the prediction of parameters. It involved several steps for designing the model that best suits the test case. 
Learning-based models have proved useful when working with one-dimensional data as input. RNN, CNN, Transformer Networks, Autoencoders, and Generative Adversarial Networks (GANs) are best suited for the task. ML algorithms such as Support Vector Machine and Random Forest Regressor were also evaluated, and they were previously quite efficient for supervised learning tasks. 

With the bloom of deep learning after AlexNet in 2012 \cite{31_NIPS2012_c399862d}, deep learning models tend to outperform the machine learning models, which also proved to be the case in our experiments. This is due to several reasons, such as their ability to automatically learn hierarchical representations of data. This means that they can extract meaningful features or representations directly from the raw data without the need for manual feature engineering. This automation is often referred to as end-to-end learning. They capture intricate patterns and generalize well, especially in scenarios where large datasets are available. Layers with non-linear activation functions allow the modelling of complex patterns and dependencies. Perhaps ML models, such as linear regression or decision trees, struggle to represent such non-linear relationships. The choice of modeling approach depends on the specific problem, the available data, the computational resources, and other factors. Different types of models and approaches may be more appropriate in certain scenarios, and it is crucial to consider the trade-offs and requirements of the problem at hand. Therefore, ML models were also tested to check their usability in this case.

\subsubsection*{Sensitivity Analysis}

The primary goal of sensitivity analysis is to understand how changes in the input parameters impact the output and to identify which inputs have the most significant influence on the model’s results \cite{32_Nusrat_2020_app10196878}. Researchers and analysts can gain insights into a model or system's robustness, reliability, and stability by conducting a sensitivity analysis.
It involved systematically changing the values of input variables within a defined range and observing the resulting changes in the model’s output or outcomes.

The training of neural networks was not so straightforward as it involved a few complexities. Therefore, a more progressive stepwise approach was adopted. (a) The data imbalance was solved through zero padding data augmentation, which made equal length for each input waveform. (b) There existed sharp and abrupt changes in some of the waveforms, which were dealt with by normalising the waveforms. Unexpected difficulties included the overfitting issue brought on by the training procedure and the model complexity. (c) A balance between the number of convolutions, pooling layers, and trainable parameters in the model was necessary to address high model complexity. (d) In order to prevent unwanted model training, early stopping was used. Different types of CNNs, RNNs and Autoencoders were experimented with to have the best model. Initially, a simple CNN was designed to check the learning process. Different factors, such as training size, noise in data, and depth of convolutions, were tested for effect. The input shape of the training batch was set as (12, 512, 1) where 12 was the batch size, 512 was the number of time stamps, and 1 was the number of bands. The number of bands was set to one since the waveform generated corresponded to the green wavelength band of the LiDAR signal. In the future, if more LiDAR data with more bands is available, it can be incorporated by changing the number of bands. Hence, adding to the modality of the network. The simulated data was stored in hdf5 format. A dataloader in Python was coded to load and split the data into train, validation, and test sets. 80:15:5 ratio was used for the split respectively.

\subsubsection*{Convolutional Neural Network}

CNN consists of convolutional layers which are the main building block of CNN to automatically learn and extract hierarchical representations of the input data. Each layer consists of one or more filters, often referred to as feature maps, that are convolved with the input data. Convolution is a mathematical operation that applies a sliding window or filters over the input data, multiplying element-wise and summing the results. Convolution helps to detect local patterns and features in the input data. After convolution, it is followed by a pooling layer often employed to reduce the spatial dimensionality while preserving the important features. This downsampling of feature maps makes the model more robust to variations and distortions in the input. Common pooling operations include max pooling, which retains the maximum value within each pooling region. Other pooling types include average pooling and min pooling. After pooling, the activation function is used typically. CNN can have linear and nonlinear activation. The non-linear activation function was one of the main breakthroughs in deep learning and resulted in the rebirth of deep learning. One such function is the Rectified Linear Unit (ReLU). It introduces non-linearities, allowing the model to learn complex relationships and capture non-linear patterns in the data. After the convolutions, two fully connected layers are employed towards the end of the network to integrate the extracted features and make predictions. These layers connect every neuron from the previous layer to every neuron in the current layer. Fully connected layers are typically followed by an activation function and a final output layer that produces the desired floating point or categorical values depending upon the last activation. Typically, the softmax activation function is used for classification tasks, while other activation functions, such as real, are used in the case of regression.

CNNs have succeeded significantly in various domains, including image classification, object detection, and image segmentation. They exploit the local correlations, spatial structure, edges, textures, and patterns. This property makes them well-suited for tasks where local details are important. CNNs are particularly effective in these tasks because of their ability to capture local features and patterns, which also have a lower number of trainable parameters compared to conventional deep networks of similar depth, making them more implementable for practical applications \cite{33_Alzubaidi_2021}.

A common myth that some people think that CNNs are limited to image-related tasks is not true as they have been proven useful in other domains, such as one-dimensional and time series data. With waveforms being 1 Dimensional, 1D convolutional layers were used to build the network. It was experimented with varying designs to come up with the best model. This includes changing a number of convolutions (convolutional depth), filters, pooling, batch normalization, encoding, decoding, and regularization. The maximum depth that could be attained with pooling after each convolution was nine due to the shape of the input signal. However, the depth was increased further with altered settings.

\subsubsection*{Recurrent Neural Network}

RNN could be highly useful for analyzing and modeling waveforms because they capture temporal dependencies and sequential information. Their success is in audio analysis, time-series forecasting, and other tasks where understanding the temporal dynamics of the data is crucial. RNNs have feedback connections that allow information to be passed from previous to current steps. This recurrent connectivity enables RNNs to capture and model sequential dependencies in the data.

Four types of RNNs are useful for dealing with sequential data including (a) Long-Short Term Memory (LSTM), (b) Bidirectional LSTM, (c) Gated Recurrent Unit (GRU) and (d) Attention Based Network. RNNs were tested using varying numbers of recurrent layers for each type of network \cite{34_WAQAS2024102946}.

\subsubsection*{Other Networks}

Apart from these, several other pretrained models were tested that were originally developed for other tasks. These include Resnet 50, 1D soat waveform, Transformers, and Wavenet. Residual Networks (ResNets) are deep CNN architectures that introduce skip connections, allowing for better gradient flow and alleviating the vanishing gradient problem. ResNets are effective in capturing both local and global features in waveform data. By enabling the network to learn residual connections, they can handle deep architectures and extract more complex representations. Transformer models, such as the ones used in natural languages processing tasks like machine translation and language modelling, can also be applied to waveform data. The self-attention mechanism in transformers allows them to capture long-range dependencies in the waveform. These models can be used for speech synthesis, music generation, and audio-to-text conversion. The BERT Transformer model was modified and tested for its performance with waveforms \cite{35_nusrat2023emojipredictiontweetsusing}. WaveNet is a deep generative model that uses dilated convolutions to model and generate waveforms. It has been specifically designed for speech synthesis and has demonstrated excellent performance in generating realistic waveforms. WaveNet can also be adapted for waveform parameter prediction tasks by modifying its output layer \cite{36_vandenoord16_ssw}. 

Deep learning models are often criticized for their lack of interpretability. One can determine if interpretability is important for the problem by comparing them with other machine learning models. If interpretability is a priority, one might choose a model that provides more transparent decision-making, such as SVM or RF. So, the performance of ML models was also compared with that of other deep learning models.

\subsection*{Real LiDAR Data}

Bathymetric LiDAR survey data was used within this research for the study areas. Initially, processing was done using the open-source library (\url{https://github.com/p-leroy/lidar\_platform}) as the data was in the las format. For visualization and processing of LiDAR data, Cloud Compare 2.11 was used in addition to the Python environment. Cloud compare provides 3D visualization of point cloud and FWL data. FWL data is computationally expensive, and it takes large storage and more time to process. Specialized processing capabilities were required for the full waveform \cite{23_Letard_2022}. The data contained following attributes: such as (a) intensity, (b) return number, (c) number of returns, (d) classification, (e) user data, (f) scan direction flag, (g) point source id, (h) gps time, (i) wavepacket index, (j) wavepacket offset, (k) wavepacket size, (l) return point wave location, (m) x (t), (n) y (t), (o)z (t), (p) XYZ, (q) depth, and (r) metadata.

The full waveform was available in a separate file in wdp format. The number of waveforms available was 30,519, while the time stamps were 168. After preprocessing, the shape of LiDAR data was (30519, 512, 1). The space required for storing these waveforms, stand-alone files, was 8 Gigabytes. For the selection of points that did not belong to the water's surface, 3D visualisation in Cloud Compare was used to segment out such points. A clear distinction between surface and sea bed points was observed. To further enhance the contrast, points were given colour on the basis of depth.

Figure \ref{fig4:3DifferenceBeweenSurfaceandSeabed} shows a visualization of real LiDAR data. The two layers are dominant in figure. The upper one shows the surface reflectance, while the lower one is the sea bed. Other points in between are reflectance from within the water column.

\begin{figure}[H]
    \centering
    \includegraphics[width=1\textwidth]{4_3DVisulizationOfDataInCloudCompare2.png}
    \caption{Difference between Surface and Sea-bed points}
    \label{fig4:3DifferenceBeweenSurfaceandSeabed}
\end{figure}

\subsection*{Attenuation Coefficient Estimation}

The attenuation coefficient of water refers to the measure of the reduction in the intensity of an electromagnetic wave as it passes through water \cite{23_Letard_2022}. It quantifies how quickly the wave's energy is absorbed, scattered, or dispersed within the water medium. The coefficient is influenced by various factors such as the wave's frequency, water temperature, salinity, and impurities or suspended particles in the water. In general, shorter wavelengths (e.g., blue or green light) tend to be absorbed more, while longer wavelengths (e.g., red or infrared light) are scattered more \cite{28_Letard9554262}. This behavior affects the penetration depth of light in water. Measuring kd provides valuable information about the strength of molecular interactions, aiding in drug discovery, understanding biological processes, and optimizing molecular interactions. Hence, it is important to know about kd.

Estimation is a tricky process as it does not vary much on a small spatial scale especially in oceans. The value of kd varies from 0.05 to 3.6 $m^-1$ in magnitude \cite{37_Corcoran:2021:0099-1112:831}. Spatial variability of kd in the study area was observed not to be changing rapidly. To get the kd value for an area, one has to do regression of intensity and depth. The log of intensity plot against depth can be seen in the following figure.

\begin{figure}[H]
    \centering
    \includegraphics[width=1\textwidth]{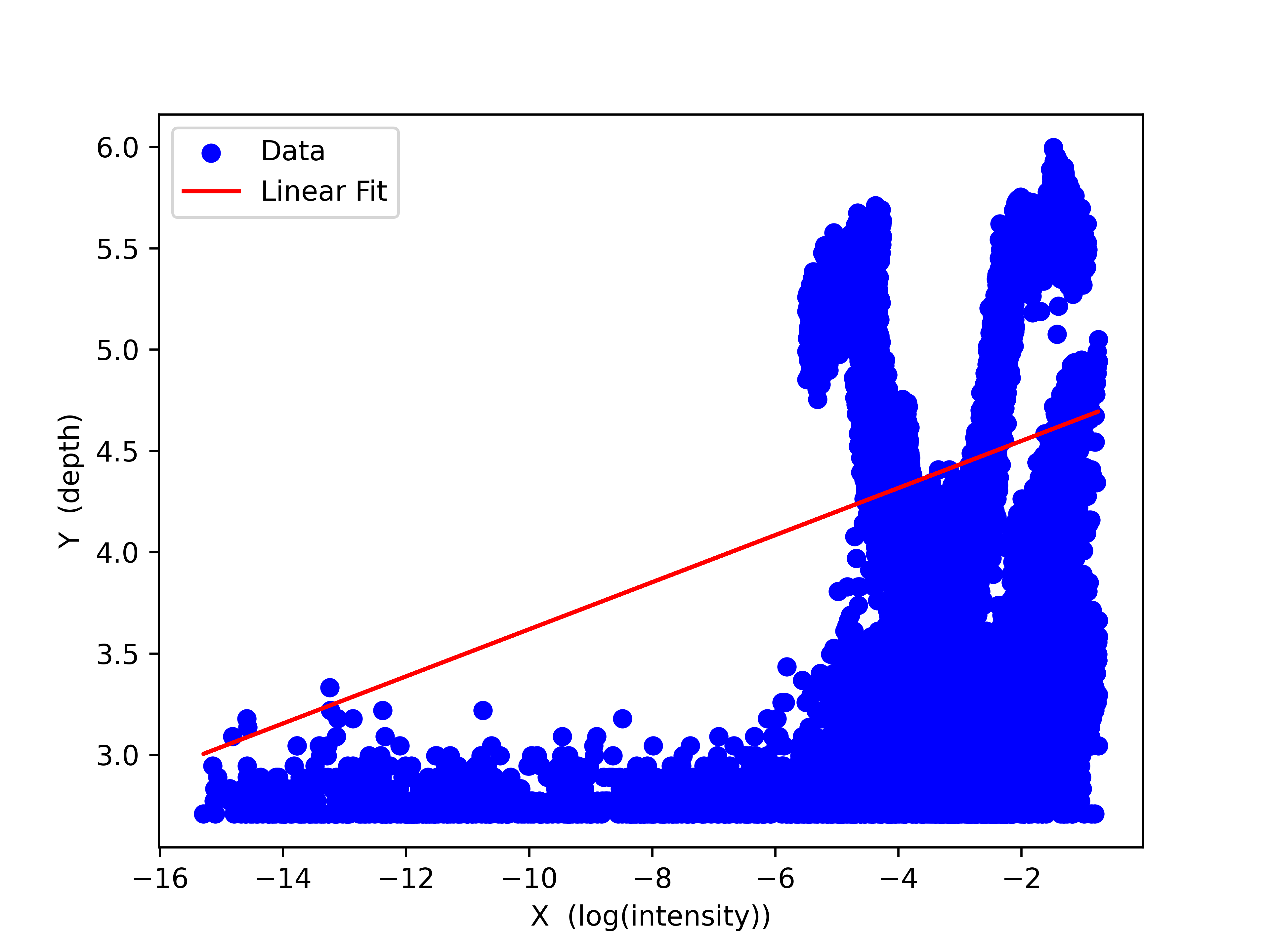}
    \caption{Attenuation coefficient assessment through regression}
    \label{fig5:KdRegression}
\end{figure}

The slope of the red line in Figure \ref{fig5:KdRegression} approximates the attenuation coefficient value. This is the only way to determine the value of kd without using deep learning. Similar regression analysis was made for 3 other real LiDAR data-sets available for having a reference kd value. It is to be noted that, the slope of the line is a single value that is a generalized value for all the data points within the plot. There is no available means, to have individual kd values for each waveform without a deep learning method. The slope of linear fit line was 0.2323 which was selected as the ground truth value for kd.

\subsection*{Domain Adaptation}

\begin{figure}[H]
    \centering
    \includegraphics[width=1\textwidth]{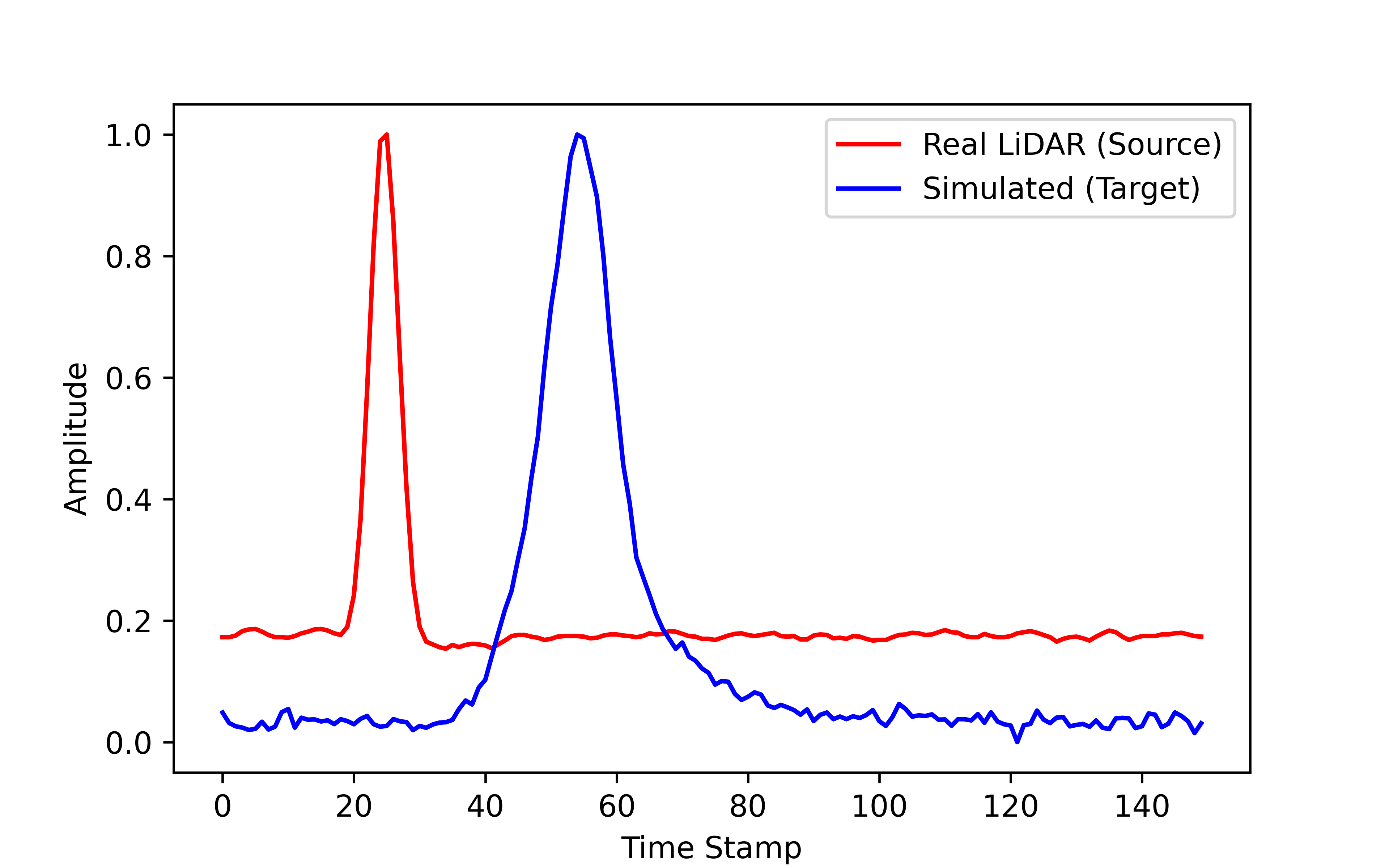}
    \caption{Difference between Waveforms}
    \label{fig6:DistibutionDifference}
\end{figure}

Domain adaptation was used to adjust the real LiDAR data in accordance with the simulated data to improve the model's performance. It is a sub-field of transfer learning to enhance a model’s performance on a target domain with insufficient annotated data by utilizing its knowledge from a related domain with sufficient labelled data \cite{38_DBLP:journals/corr/abs-2010-03978}. It is a commonly used technique in RS, especially concerning satellite imagery. Satellite images of different places show different characteristics. For example, satellite images from Pakistan will differ from those in Austria or France. Therefore, for any task such as land use classification, a model that is trained on one might not perform well enough on the other. Figure \ref{fig6:DistibutionDifference} shows a plot of a sample of normalized waveforms, which led to the decision to carry out domain adaptation of the data.

For the second test of our model, the trained model was used to predict labels for another simulated dataset generated using a different simulator. The model could predict the labels, but without any further training, the results were not good enough. The model improved significantly with slight training on the other data, leading to better predictions. The coefficient of determination, r2 score, jumped up to 0.49.

Domain adaptation techniques were explored for LiDAR datasets generated from different simulators. Multiple domain adaptation techniques were tested, including data adaptation and model adaptation. For data adaptation, optimal transport was used, as suggested by Courty et al. \cite{39_Courty_758603}.
One of the main challenges of domain adaptation is that labeled data may not be available in the target domain. This makes it difficult to train a classifier that can generalize well to new data in that domain. Unsupervised domain adaptation is when labeled data is unavailable in the target domain. This makes it difficult to train a classifier that can generalize well to new data in that domain. A common feature representation for both domains, latent space, was found to address this challenge. This allows labelled samples from the source domain to train a classifier that can also be applied to the target domain. Courty et al. \cite{39_Courty_758603} proposed a regularized unsupervised optimal transportation model to align these representations and exploit both labelled samples and distributions observed in both domains. Thus, deriving the source and target domains' probability density functions (PDFs). Among the unsupervised domain adaptation, sinkhorn transport worked better.

\subsubsection*{Optimal Transport}

Optimal Transport is a mathematical function that quantifies the optimal way to redistribute mass or resources from one distribution to another. It provides a principled approach to compare and measure the dissimilarity between probability distributions. In the context of transportation, optimal transport seeks to find the most efficient way to move mass from a set of source points to a set of target points such as to minimise the total cost required for the transportation.

A transportation plan often represents the concept of optimal transport. A transportation plan is a matrix that specifies how much mass from each point in the source domain should be transported to each point in the target domain. The process of learning a transportation plan involves minimizing an objective function based on optimal transport, which is regularized to ensure that labeled samples of the same class in the source domain remain nearby during the transportation process. The researchers provided a python package for optimal transport for experimenting with types of optimal transport \cite{39_Courty_758603}. The tested types were EMD Transport, Sinkhorn Transport, and Sinkhorn Transport with regularization.

\section*{Results and Discussion}

This section presents a comprehensive analysis of the findings obtained from the conducted study. It aims to provide a detailed examination and interpretation of the data, shedding light on the key outcomes and their implications. The study focused on the development of simulated data, the design of a suitable model and the estimation of parameters. In this section, obtained results are described and discussed stepwise. Furthermore, we delve into potential explanations for observed patterns and variations, considering both anticipated and unexpected outcomes. Through a rigorous analysis and contextualization of the data, this section provides insights into the implications of the study findings, their relevance to the research objectives, and their broader implications in the field.

The first task was to check the effect of input parameters for the simulator. An effort was made to simulate waveforms as similar as the Titan sensor waveforms. The approach was successful to a good extent by using the correct parameters. Table \ref{tab1:InputParameters} describes the range of parameters used in the WALID Simulator.

\begin{table}[H]
    \centering
    \begin{tabular}{c|c|c}
    \hline
    \textbf{Parameter} & \textbf{Low Limit} & \textbf{High Limit} \\
    \hline
         Depth & 0.15 & 19  \\
    
         $I_{ref}$ & 1 & 100 \\
         kd & 0 & 1 \\
         $I_w$ & 0 & 2 \\
         A & 1 & 10 \\
         noise & 0 & 0.04*A \\
         Imp Type & 0 & 2 \\
         $W_c$ & 0.1 & 1 \\
         \hline
         
          
    \end{tabular}
    \caption{Input Parameters for WALID Simulator}
    \label{tab1:InputParameters}
\end{table}

\subsection*{Model’s Design}

An analysis was performed to derive the effect of different parameters on the model performance. These parameters included batch size, train-validation split ratio, choice of the optimizer, and learning rate. Other factors were examined, such as the effect of deriving noise on data, activation functions, kernel size, number of trainable parameters, feature maps, and convolution depth for the sensitivity analysis.
Thus, this step helped in decision-making for selecting a suitable model.

\subsubsection*{Effect of Noise}

Adding sufficient noise to the data made it more similar to the actual data. It improved the model performance by making the training process more robust. Such was the case with the experiment. Adding a small amount of noise to the data improved the model performance often as it also happened in other studies \cite{40_Noise_10.1162/neco.1995.7.1.108}. Injecting noise into the input to a neural network can also be seen as a form of data augmentation.

\begin{figure}[H]
    \centering
    \includegraphics[width=0.75\textwidth]{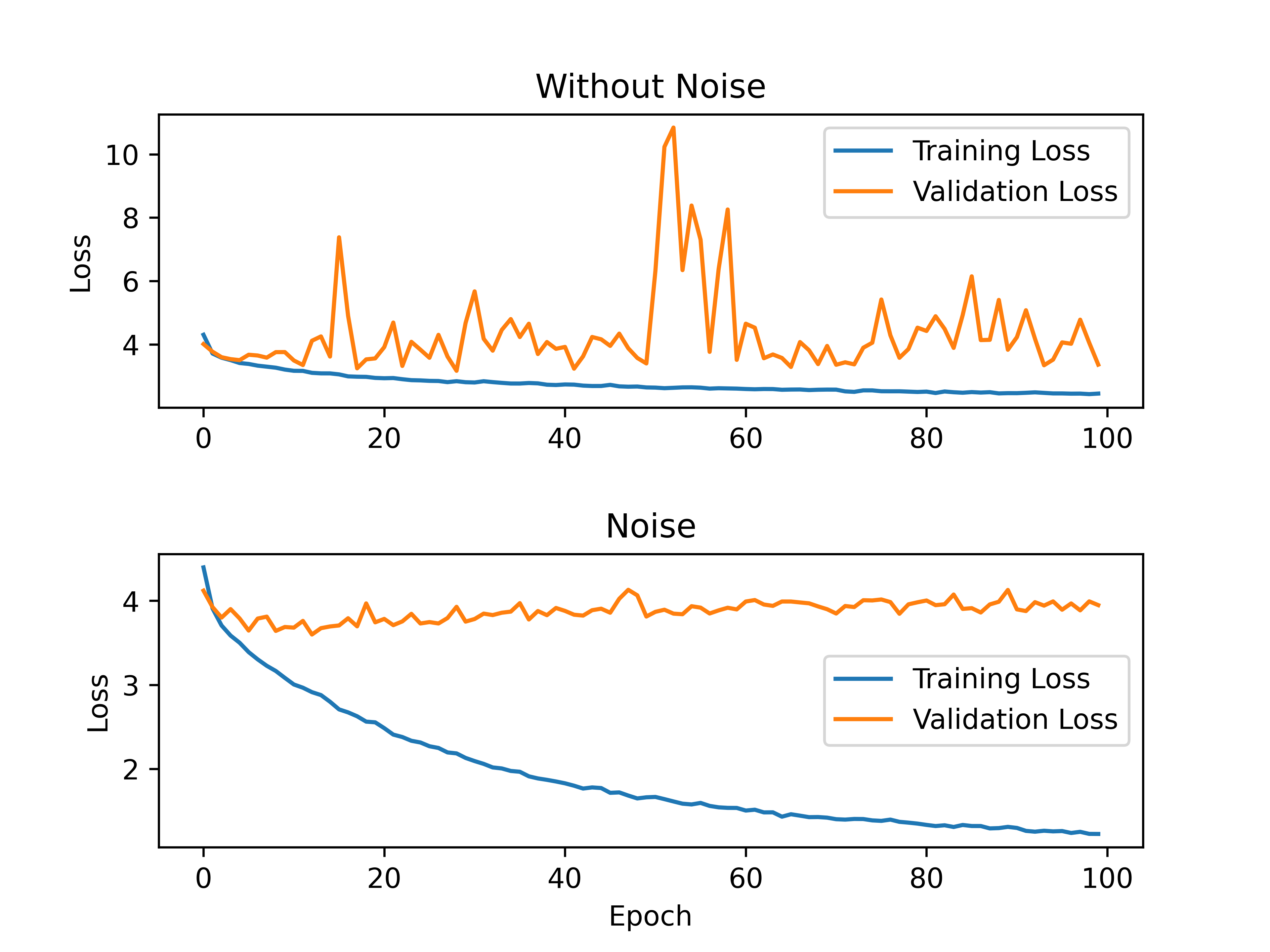}
    \caption{A comparison showing the effect of noise on training process}
    \label{fig7:EffectofNoise}
\end{figure}

A comparison is available in Figure \ref{fig7:EffectofNoise}. The noise was added through a parameter value in the waveform simulator. Visual differences in a signal with and without noise can be seen in the figure. With noise, the model can learn the noise to better predict on unseen data, which might be contrary to what one might think initially. In Figure \ref{fig7:EffectofNoise}, large fluctuations in validation loss can be observed because the model cannot generalize well on unseen data. With noise in training data, smoother learning curve with gradual change was obtained.

\subsubsection*{Effect of Convolutional Depth}

An increase in convolutional depth up to a certain number increased the model performance, whereas any further increase resulted in overfitting with many trainable parameters. 

\begin{figure}[H]
    \centering
    \includegraphics[width=0.75\textwidth]{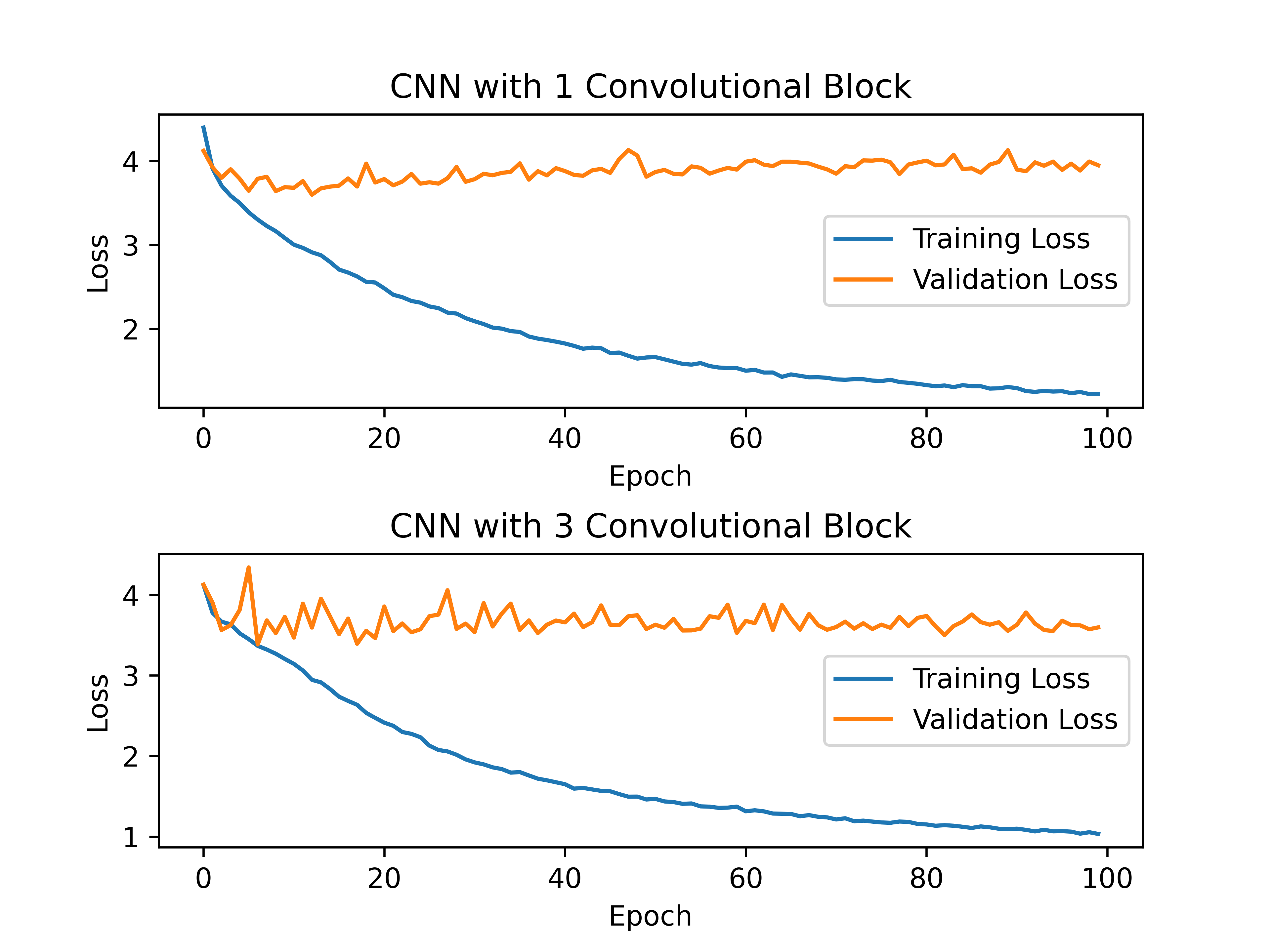}
    \caption{A comparison showing the effect of Convolutional Depth on training process}
    \label{fig8:EffectOfConvolutionalDepth}
\end{figure}

With fewer convolutions, the model converged faster. This was because it did not have enough trainable parameters with a smaller number of convolutional layers. This effect is shown in Figure \ref{fig8:EffectOfConvolutionalDepth} as observed during the training.

\subsubsection*{Effect of Loss Function}

The loss function quantifies the discrepancy between the predicted and actual outputs, and then the model tries to minimize this loss during training. The loss function affects how the model parameters are updated after each batch. Different loss functions emphasize different aspects of the training process, such as accuracy, robustness to outliers, or handling class imbalance RMSE, MAE, Huber and Log-Cosh loss were tested. Then, MAE was selected as the loss function for further use. This resulted in lower loss and indicated the magnitude of error in the model predictions as it is more interpretable. MAE measures the average absolute difference between the predicted and true values and is also less susceptible to outliers, which was desirable in our case. MAE is given by the equation \ref{eq5} where y, $\hat{y}$ and n	is the	prediction, true value and number of points respectively.
\begin{equation} \label{eq5}
\text{MAE}(y, \hat{y}) = \frac{ \sum_{i=0}^{N - 1} |y_i - \hat{y}_i| }{N}
\end{equation}

\subsection*{Model’s Performance}

The performance of models was tested through a series of steps. Initially, models' ability to learn through data was known through observing training and testing loss. The second evaluation was to test the model's prediction on unseen data. A test set with waveform samples generated through the same simulator was used to evaluate model performance. Then, another test set generated from another LiDAR simulator with different features was used to further evaluate the model's performance. Finally, real LiDAR data from a bathymetric survey in the Northern coastal area of Brittany was used to evaluate the model. The results of the model’s performance on unseen simulated data are discussed below. The following tables present model performance for predicting the depth of water. Each better-performing model type was tested to predict the other two parameters. The same model did not have the same performance for each parameter. Though the input is the same for predicting each parameter, learning for different parameters is required for predicting different parameters. 

\textbf{\textit{Convolutional Neural Networks (CNN):}} 1D CNN showed the best results for predicting the three parameters. Multiple experiments were performed by changing the convolutional depth, pooling and filters. 

\textbf{\textit{Recurrent Neural Networks (RNN)}:} RNN were supposed to be more relevant for dealing with sequential data, yet CNN outclassed slightly for this task. RNN models, though not the best, still showed promising results. Among the RNN models, Attention based models were the most significant in terms of performance metrics.

\textbf{\textit{Machine Learning (ML) Models:}} ML models could not predict the parameters to an acceptable level. Interpretability was not a priority in cases where the ML model was not able to provide comparable results. Table \ref{tab2:CNNRNNOthersTable} also shows the performance of ML models.

\textbf{\textit{Pretrained Models:}} The pretrained models (Resnet 50, 1D soat waveform, Transformers, and Wavenet) were fine-tuned on the simulated data. However, none of the models, that were tested, were able to improve significantly on already-used models. 
Table \ref{tab2:CNNRNNOthersTable} illustrates the performance of CNN, RNN, ML and Pretrained models for predicting depth parameters. In the following tables, the $r^2$ score (t), $r^2$ score (v) and $r^2$ score represents coefficient of determination during training, validation and testing phase. The dashed cells show results that were not trivial.

\begin{table}[H]
    \centering
    \begin{tabular}{@{}p{2cm}p{1.5cm}p{1.5cm}p{1.5cm}p{1.5cm}p{1.5cm}p{1.5cm}p{1cm}@{}}
    \hline\hline
        \textbf{Type} & \textbf{Training Loss} & \textbf{Validation Loss} & \textbf{$r^2$ score} (t) & \textbf{$r^2$ score} (v) & \textbf{$r^2$ score} & \textbf{RMSE} \\
    \hline
Conv1D 1 & 2.209   &	4.479	& - & - &	- &	-  \\
-- 1D 3 & 1.041	& 3.6864    & -	& -	& -	& - \\
-- 1D 5   & \textbf{0.613}	& 3.95     	& \textbf{0.923} &	0.314 &	0.479 &	1.502 \\
-- 1D 5+2D & 2.284	& 2.522     &	0.591 &	0.582 &	0.577	\\
-- 1D 9 & 1.296	& 3.25	& - &	- &	-	& - \\
-- 20 Up & 2.661	& 3.270 &	0.782	& 0.184	& 0.175	& 2.531 \\
-- 20 Down & 1.096 &	\textbf{2.120}	& 0.514	& 0.310	& 0.302	& 3.785 \\
\hline

    \hline
        RNN LSTM 1 & 4.7868	& 4.7451	& -	& -	& -	& - \\
        LSTM 4 & 3.6405	& 3.606	& 0.0025	& 0.002	& 0.002 & - \\
        Bidirectional &  4.783 & 4.7467 & 0.0021  & 0.0021 & 0.0028 & -  \\
        GRU & 4.7088	& 4.6288	& 0.0002	& 0.0001	& 0.007	& -  \\
        Attention Based & 3.114	& 3.134	& 0.402	& \textbf{0.383}	& \textbf{0.35}	& 4.199 \\
    
    \hline

SVM Linear	& -	& -	& 0.715	& 4e-10	& 347e-10	&- \\
SVM Polynomial	& -	& -	&-	& 9e-10	& -	& - \\ 
SVM rbf	&-	&-	& 1973e-10	& 8e-7	& -	& - \\ 
SVM sigmoid	&-	&-	&-	& 6e-10	& 6.071e-7	&- \\ 
Random Forest	&-	&-	&-	&3.975e-7	& -	& - \\
\hline

Resnet-50	& 3.064	& 3.169	& 0.399	& 0.350	& 0.296	& 3.930 \\
1D soat waveform	& 0.973	& 3.745	& 0.754	& 0.175	& 0.124	& - \\
Transformers	& 0.87	& 3.2	& 0.896	& 0.351	& 0.381	& - \\
Wavenet	& -	& -	& -	& 2.15E-11	& 4.15E-15	&- \\
\hline
\hline

    \end{tabular}
    \caption{Performance of Convolutional Neural Network, Recurrent Neural Networks, Machine Learning models and Pretrained Models}
    \label{tab2:CNNRNNOthersTable}
\end{table}

\subsubsection*{Model with Best Performance}

With rigorous testing, a CNN model was selected as best performing model. There were three methods identified to compile the final model to predict three parameters \cite{41_jamal2023data}. The first was to have a single network with three filters in the last layer to predict three outputs; the other way was to to jointly learn feature maps and three fully connected layers. The third approach was to have three separate branches for three types of predictions with same input. All three approaches were tested and the network with three branches performed the best and was selected as the final model.

\begin{figure}[H]
    \centering
    \includegraphics[width=1\textwidth]{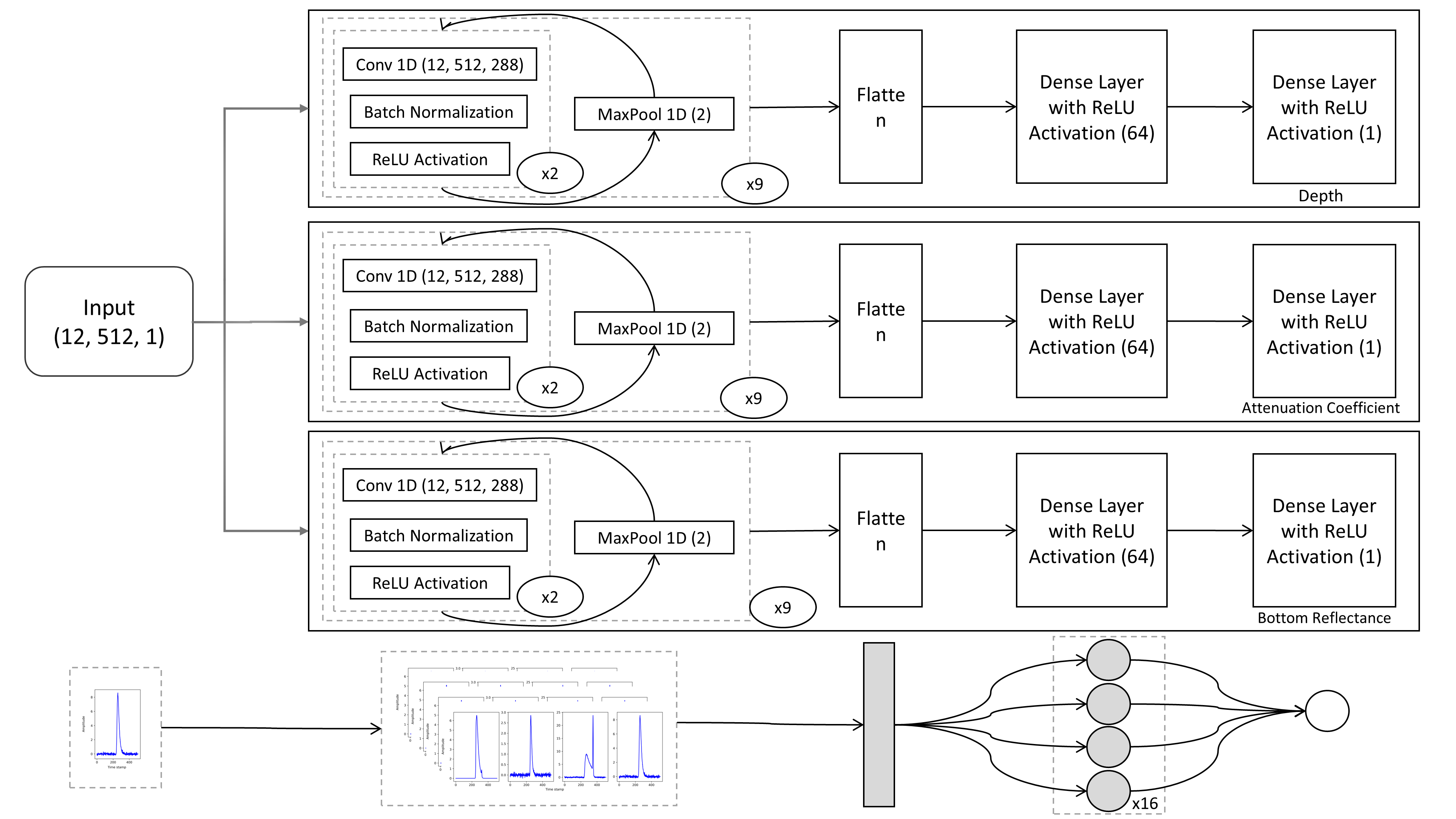}
    \caption{Model summary and conceptual sketch}
    \label{fig9:ModelSummary}
\end{figure}

The model had three branches for predicting three parameters. The three branches had the same input: waveforms. The three branches were compiled into a single giant model. This was the final model, which had a total number of parameters 5,068,233, trainable parameters of 5,049,993, and non-trainable parameters of 18,240. A convolution consists of a 1D convolutional layer, a batch normalization layer and an activation layer with a ReLU function. Each branch had eighteen convolutions with a Max Pooling layer after every second convolution. After the convolutional layers, fully connected dense layers were used after flattening. Figure \ref{fig9:ModelSummary}  presents an abstraction of the model waveform processed through the neural network.

The following table provides the performance of the final model after testing through unseen data. This means the model’s effectiveness and accuracy were assessed on data it had not been exposed to during its training or validation stages. The table offers insights into how well the model generalizes to new, unseen data, comprehensively assessing its performance in real-world scenarios. The best performance model had the lowest error and highest $r^2$ score, on the test data. Tables \ref{tab3:best_performing_model_table}, \ref{tab4:PerfomanceOtherSimulatedData} and \ref{tab5:PerfomanceRealData} present the metrics for the best-performing model, the test data from the same simulator, test data from the other simulator and test data from real LiDAR, respectively.

\begin{table}[H]
    \centering
    \begin{tabular}{@{}p{2cm}p{1.5cm}p{1.5cm}p{1.5cm}p{1.5cm}p{1.5cm}p{1.5cm}@{}}
    \hline
\textbf{Parameter}	& \textbf{Training}	& \textbf{Validation}	& \textbf{Training $r^2$ score}	& \textbf{Validation $r^2$ score}	& \textbf{Test $r^2$ score} & \textbf{RMSE} \\
 \hline
depth	& 2.06	    & 2.069	& 0.6617	& 0.6489	& 0.6518 & 3.185 \\
kd	    & \textbf{0.0252 }	& \textbf{0.0264}	& \textbf{0.9565}	&   \textbf{0.9541} & \textbf{0.9506} & \textbf{0.0644}\\
bottom	& 16.74	& 16.82	& 0.389	& 0.374	& 0.367 & 22.76 \\

\hline

    \end{tabular}
    \caption{Performance of best performing model on test data}
    \label{tab3:best_performing_model_table}
\end{table}


\begin{table}[H]
    \centering
    \begin{tabular}{@{}p{5cm}p{1.5cm}p{1.5cm}@{}}
    \hline
        \textbf{Type} &  \textbf{MAE} & \textbf{RMSE}  \\
    \hline
\textbf{depth} \\
trained network	& 	  3.895 	& 4.69	\\
domain adaptation & 5.485 & 6.486 \\
fine tuning &  \textbf{1.006}  & \textbf{1.4514} \\

\textbf{kd} \\
trained network	& 	 0.2411  	& 0.2949	\\
domain adaptation & 0.2718 &  0.3366 \\
fine tuning &  0.2911  &  0.3395 \\

\textbf{bottom} \\
trained network	& 	32.03  	&  34.93	\\
domain adaptation & 46.031 &  46.031 \\

\hline
 
    \end{tabular}
    \caption{Performance of Best performing model on other simulated data}
    \label{tab4:PerfomanceOtherSimulatedData}
\end{table}


\begin{table}[H]
    \centering
    \begin{tabular}{@{}p{5cm}p{1.5cm}p{1.5cm}@{}}
    \hline
        \textbf{Type} &  \textbf{MAE} & \textbf{RMSE}  \\
    \hline

\textbf{depth} \\
trained network	& 	  4.49 	& 4.9	\\
domain adaptation & 2.319 & 2.779\\
fine tuning	 &  \textbf{1.616}	& 2.1	 \\

\textbf{kd} \\
trained network	& 	  0.3674 	& 0.3713	\\
domain adaptation &  \textbf{0.0358} & 0.0402 \\
fine tuning	 &  0.0358	& 	0.0358 \\

\hline
 
    \end{tabular}
    \caption{Performance of best performing model on real data}
    \label{tab5:PerfomanceRealData}
\end{table}

The model exhibited favorable performance when applied to previously unseen data. Model performance was excellent for the test-simulated data on which it was trained. A slight decrease in performance was observed when predictions were made on simulated data and real LiDAR data. Therefore, techniques such as domain adaptation and transfer learning were used. Different domain adaptation and transfer learning techniques were experimented with. The results above pertain to the better-performing models. The model's performance increased significantly on real LiDAR data after domain adaptation and fine-tuning.

\section*{Conclusion}

A one-dimensional Convolution Neural Network successfully produced a data-driven approximation of the implicit relation between waveform and parameters. This model is the first of its kind; it will enhance the usability of Full-Waveform LiDAR (FWF) by adding important parameters to bathymetry through LiDAR. A total of one million simulated waveforms generated using the simulator were used to train the deep learning model. 

The proposed model showed exceptionally high performance in predicting kd with a correlation coefficient of 0.95 and a loss of 0.028 for test data. The model made significant good predictions for depth parameters. However, the model could not show a high correlation for bottom reflectance, in which there is further room for improvement. Bottom reflectance is one of the most challenging parameters to predict since little or no clear signal can be observed in most cases. Nevertheless, predictions made by the model are still useful for estimating bottom reflectance. 

Future perspectives include creating feature space using real LiDAR data with an autoencoder network and then using optimal transport with the feature space instead of transferring real LiDAR data to the original waveform space. This technique tends to work better in most cases. With the availability of more data in the future, more robust models can be trained with no further need for domain adaptation for data with different characteristics.

\section*{Authorship contribution}
S.A.J. – Methodology, Material preparation, Investigation, Formal analysis, Software, Visualization, Writing – original draft; T.C. – Conceptualization, Supervision; D.T. – Supervision, Resources, Validation; M.L. – Conceptualization, Material preparation, Methodology, Validation, Supervision; D.L. – Data Curation; 
\section*{Conflicts of Interest }
     The authors declare no conflict of interest.
\section*{Acknowledgement}
The author acknowledge the supervision, conceptualisation of topic and discussions from Dr. Thomas Corpetti, Dr. Mathilde Letard, Dr. Dimitri Lague and Dr. Dirk Tiede. 
The author received a European scholarship to engage in Master Copernicus in Digital Earth, an Erasmus Mundus Joint Master Degree (EMJMD). 

\section*{Data Availability}
The datasets generated and analysed during the current study along with the model and code are available in the github repository:  \url{https://github.com/SaadAhmedJamal/LiDAR}.

\bibliography{sample}

\end{document}